\documentclass[prb,twocolumn,showpacs,floatfix,superscriptaddress,amsmath,amssymb]{revtex4}
\usepackage{graphicx}
\usepackage{graphics}
\usepackage{color}
\usepackage{pst-all}

\newcommand{\beq}{\begin{equation}}
\newcommand{\eeq}{\end{equation}}
\newcommand{\beqa}{\begin{eqnarray}}
\newcommand{\eeqa}{\end{eqnarray}}
\newcommand{\nn}{\nonumber}

\begin{document}

\title{Coexisting orders in the quarter-filled Hubbard chain with elastic deformations}
\author{H.D.\ Rosales}
\affiliation{Departamento de F\'{\i}sica, Universidad Nacional de la Plata, C.C.\ 67, (1900) La Plata, Argentina}
\author{D.C.\ Cabra}
\affiliation{Departamento de F\'{\i}sica, Universidad Nacional de la Plata, C.C.\ 67, (1900) La Plata, Argentina}
\affiliation{Institut de Physique et Chimie des Mat\`eriaux de Strasbourg, UMR 7504, CNRS-UdS,
23 rue du Loess, BP 43, 67034 Strasbourg Cedex 2, France}
\affiliation{Facultad de Ingenier\'ia, Universidad
Nacional de Lomas de Zamora, Cno.\ de Cintura y Juan XXIII, (1832)
Lomas de Zamora, Argentina.}
\date{\today}

\begin{abstract}
The electronic properties of the quarter-filled extended Peierls-Holstein-Hubbard
model that includes lattice distortions and molecular deformations are investigated
theoretically using the bosonization approach. We predict the existence of a wide
variety of charge-elastic phases depending of the values of the Peierls and Holstein
couplings. We include the effect of the Peierls deformation in the nearest-neighbor repulsion $V$, that may be present in real materials where Coulomb interactions depend strongly on the distance, and we show that the phase diagram changes substantially for large $V$ when this term is taken into account.
\end{abstract}
\maketitle

\section{Introduction}
The electronic properties of the low dimensional interacting electron systems have attracted great interest for many different reasons, the main being that they are simpler to analyze than higher dimensional ones and they could then be used to gain some insight on their higher dimensional counterparts. Typically this connection can be envisaged by coupling many 1D systems to build up a higher dimensional array. They are also interesting on their own since many materials show a quasi-one-dimensional behavior in a certain range of temperatures where
higher dimensional couplings can be neglected. A third motivation to study 1D systems comes from the very exciting cold atom systems which provide a test laboratory for many important theoretical developments already done and also motivates further analysis that could be in principle directly tested in real systems.\\
In particular, and in connection with real materials, quarter-filled systems have received a lot
of attention since they are very good candidates to describe the so called organics charge transfer salts like the Bechgaard salts (TMTSF)$_2X$ or (TMTTF)$_2X$ whre X=PF$_6$, AsF$_6$\cite{Salts}. The Hubbard model is the basic model for understanding the electronic properties of these quasi 1-D organic conductors and in some cases, lattice deformations could play a central role. In addition to this, in crystalline materials where the building block of the crystal structure is a large molecule, the vibrational  properties of the molecules often have large effects on the electronic properties of the material\cite{Salts}. These materials exhibit a variety of electronic states with spatially inhomogeneous charge, spin and lattice structures.

Several experimental studies have shown clear evidence for the existence of charge order
\cite{Hiraki98, Mazumdar99B, Hiraki99,Biskup,Chow,Nad00A,Nad00B,Monceau,Miyagawa}
in organic charge transfer solids and this has stimulated considerable theoretical efforts
\cite{Ung,Mazumdar99A,Seo97,Seo00,Kobayashi97,Kobayashi98,Riera99,Riera00}.
These materials involve Coulomb repulsion, both on-site
and between nearest-neighbors ($U$, $V$ respectively) and are $1/4$-filled. Due to the nearest-neighbors Coulomb repulsion, it is expected that in these materials a charge ordering  $\bullet-{\small\text{o}}-\bullet-{\small\text{o}}$ sets in, where $\bullet$ and ${\small\text{o}}$
correspond to occupied and unoccupied sites. The charge order $\bullet-{\small\text{o}}-\bullet-{\small\text{o}}$
corresponds to the so called $4k_F$ Charge-Density-Wave (CDW), where $k_F = \pi/4\,a$ with $a$ the lattice constant.
However, for large  values of $V$, and when electron-phonon interactions are included,
the ground state corresponds to an ordering like $\bullet-\bullet-{\small\text{o}}-{\small\text{o}}$
\cite{Ung,Mazumdar99A,Kobayashi97,Kobayashi98,Riera99,Riera00}.
The charge order corresponds to a tetramerized
$2k_F$ CDW and it coexists with the $2k_F$ periodic modulation of the intersite distances, called Bond-Ordered-Wave (BOW) or a mixed $2k_F$ + $4k_F$ BOW. This charge order
is called Bond-Charge Density Wave (BCDW). On the other hand, if the Coulomb interaction between two sites $r_i$ and  $r_j$, is exponentially short
ranged, {\it i.e.} of the form $\exp[-\alpha|r_i-r_j|]/r^\xi$ with $\alpha,\xi>0$, lattice distortions may have
an important effect on the Coulomb interactions between nearest-neighbor.

Motivated by the previous discussion we study, using the Abelian bosonization approach and a semiclassical
analysis of the effective theory, the ground state phase diagram of the Peierls-Holstein-Hubbard model including the effects of the Peierls deformation in the nearest-neighbor Coulomb interaction $V$.
We show that the ground state at zero temperature indeed favors several period two and period four
distortion patterns, stemming from a competition between elastic  and charge-spin
energy. These patterns spontaneously break translational symmetry, with different phases
depending on the  Coulomb interaction strengths $U$ and $V$ and on the value of the spin-phonon couplings.
A rich phase diagram is obtained, including all the combinations of Charge Order and $2k_F-4k_F$
deformations as a function of the spin-phonon couplings.

When the effect of the Peierls deformation in the nearest-neighbor Coulomb interaction is not considered, our analysis reproduces previous results obtained with numerical methods\cite{Clay03} .
The main novelty in our treatment is the inclusion of a nearest neighbors Coulomb repulsion $V$ which
depends on the distance between adjacent ions and hence depends on the lattice distortions. The phase diagram
then changes substantially indicating that this variation on $V$ should not be neglected in comparing
to experiments. Our analysis allows for a description only at a qualitative level, but it has the advantage of
being easily generalizable to include other perturbations as longer range Coulomb interactions,
to treat arbitrary filling fractions, to add an external magnetic field, etc.

The paper is organized as follows. In Section II we present the model and its analytical treatment.
The phonon sector is described in the adiabatic approximation by classical static deformations
and we analyze the possible patterns. In Section III, we use the bosonization approach to represent
the electrons operators in terms of bosons and analyze the effective description by considering
all the relevant perturbation terms as semiclassical potentials. We draw a qualitative phase
diagram with from analisys. Special emphasis is put on the characterization
of the ground state phases that result from the combination of charge-spin degrees of freedom
and elastic effects in different parameter ranges. Finally, in Section IV we present a summary
and conclusions of the present work.
\section{Model Hamiltonian}

We consider a $1D$ extended Hubbard  model with Hamiltonian given by

\begin{eqnarray}
\label{H_total1}
H &=&-\sum_{j,\sigma}t_{j,j+1}[c_{j,\sigma}^\dagger c_{j+1,\sigma}+
c_{j+1,\sigma}^\dagger c_{j,\sigma}]+ \nn\\
&&+ U\sum_{j}n_{j,\uparrow}n_{j,\downarrow} \nn +
\sum_{j}\,V_{j,j+1}\,n_{j}n_{j+1}+ \nn\\
&&+\sum_{j}\beta_j\,n_{j}
\end{eqnarray}
where $c_{j,\sigma}$ denotes the annihilation operator of an electron with spin
$\sigma (=\uparrow, \downarrow)$ at the $j$th site, and $n_j=n_{j,\uparrow}+n_{j,\downarrow}$
is the charge density with $n_{j,\sigma}=c_{j,\sigma}^\dagger c_{j,\sigma}^{}$.
$t_{i,i+1}$ is the hopping integral,  the parameters  $U(>0)$ and $V_{j,j+1}(>0)$ denote the
magnitudes of the on-site and nearest neighbor interactions and $\beta_j$  represents the
internal molecular deformation. We focus on the $\frac{1}{4}$-filled case, with the average number
of electrons per site $\langle n_i\rangle=1/2$.

The interaction of the charge degrees of freedom  in a homogeneous Hubbard chain
($t_{j,j+1}=t$) with phonons is usually
modeled by a linear expansion of the exchange couplings around the non distorted values $t$,
\begin{eqnarray}
t_{j,j+1}&\simeq&t(1-g_P\,\delta_{P,j})\nn
\end{eqnarray}
where $\delta_{P,j}$ is the local lattice distortion and $g_P$ is the electron-phonon interaction.
On the other hand, local deformations of the molecules can produce changes of the on-site (or molecular)
orbital energies and can simply be taken into account by a Holstein term
\begin{eqnarray}
g_H\sum_{j}\delta_{H,j}n_{j},
\end{eqnarray}
where  $g_H$ is the electron-phonon coupling constant of the on-site type  (in Eq.\ (\ref{H_total1}) we change $\beta_j\rightarrow g_H\,\delta_{H,j}$).

In addition to the purely electronic terms in  Eq.\ (\ref{H_total1}) and unlike previous analysis, we
include  the effects of the Peierls deformation in the nearest-neighbor Coulomb interaction $V$ that
may be important when the Coulomb interaction between two sites $r_i$ and  $r_j$ is exponentially short ranged, {\it i.e.}
$\exp[-\alpha|r_i-r_j|]/r^\xi$. Specifically, we perform a linear expansion of the $V_{j,j+1}$ couplings around the non
distorted values $V$,
\begin{eqnarray}
V_{j,j+1}&\simeq&V(1-g_V\,\delta_{P,j})\nn
\end{eqnarray}
where $g_V$ is the coupling constant which measures the effects of the bond deformation on the Coulomb repulsion.\\
The complete Hamiltonian, including  the elastic energy in the adiabatic approximation, is written as
\begin{eqnarray}
\label{H_total}
H &=&-t\sum_{j,\sigma}(1-g_P\,\delta_{P,j})[c_{j,\sigma}^\dagger c_{j+1,\sigma}+
c_{j+1,\sigma}^\dagger c_{j,\sigma}]+ \nn\\
&&+ U\sum_{j}n_{j,\uparrow}n_{j,\downarrow} \nn +
V\sum_{j}(1-g_V\,\delta_{P,j})\ n_{j}n_{j+1}+ \nn\\
&&+ g_H\sum_{j}\delta_{H,j}n_{j} \nonumber \\
&&+ \frac{K_P}{2}\sum_{j}\delta_{P,j}^2 + \frac{K_H}{2} \sum_{j}\delta_{H,j}^2
\end{eqnarray}
where the parameters $K_H$ and $K_P$ are the conventional elastic constants for the on site deformation
and the lattice distortion, respectively.

\begin{figure}[tbp]
\centering
\includegraphics[width=0.41\textwidth]{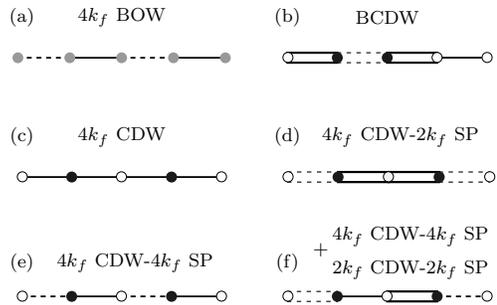}
 \caption{Possible lattice deformations and charge density wave ground state in the
Peierls-Holstein-Hubbard model at quarter filling. The Grey, black and white circles correspond
to site charges of $\langle n\rangle=$ $0.5$, $1$ and $0$, respectively. We reproduce the previous results of \cite{Clay03} and we
show two new phases $(e)$, $(f)$ that appear when $g_V > 0$. In this figure (a) The $4k_F$ BOW state,
with dimerized bond orders and uniform site charge densities, in (b) represent the
BCDW state, with bond orders $2k_F>4k_F$ and  charge order
$\bullet-\bullet-{\small\text{o}}-{\small\text{o}}$. (c) The $4k_F$ CDW
with $\bullet-{\small\text{o}}-\bullet-{\small\text{o}}$  charge order and uniform bond order.
(d) The $4k_F$ CDW-$2k_F$ SP state, with $\bullet-{\small\text{o}}-\bullet-{\small\text{o}}$
charge order and bond orders $2k_F$. For $g_V = 0.2$, we have phase (e) {\it i.e.}, $4k_F$ CDW-$4k_F$ SP with $\bullet-{\small\text{o}}-\bullet-{\small\text{o}}$
charge order together with dimerization bond order. Finally, in the phase (f) ($4k_F$ CDW-$4k_F$ SP + $2k_F$ CDW-$2k_F$ SP) is characterized by three different site charges $\frac{1}{2}-\varepsilon'$, $\frac{1}{2}+\varepsilon$, $\frac{1}{2}-\varepsilon'$,
$\frac{1}{2}+\varepsilon''$  with bond orders $2k_F>4k_F$.}
\label{pattern}
\end{figure}

Now, we consider the lattice deformations. At quarter-filled ($\langle n_i\rangle=1/2$) the most general period four lattice deformations without collective displacement, can be parametrized as
\begin{eqnarray}
\delta_{P,j}&=&\delta_{P,d}\cos(\pi\,x_j)+\delta_{P,t}\cos(\pi/2\,x_j+\xi-\pi/4)\nn\\
\delta_{H,j}&=&\delta_{H,d}\cos(\pi\,x_j)+\delta_{H,t}\cos(\pi/2\,x_j-\zeta)
\label{deformations}
\end{eqnarray}
where $x_j$ is the position of the $j-th$ site (lattice constant $a$ has been set to one),
and the phases  $\xi$ and $\zeta$ determine the spatial patterns of tetramerization
$\delta_{P,t}$ and $\delta_{H,t}$ respectively, while $\delta_{P,d}$ and $\delta_{H,d}$
are the amplitude of the  dimerization  in the lattice distortion and the intrasite electron-phonon deformation. This parametrization is the most general supported by bosonization, as such a deformation is commensurate. Period four deformations cause commensurability of relevant perturbations at $\langle n_i\rangle=1/2$ and provide a mechanism for a charge gap in this regime\cite{Daniel01}. Numerical evidence of the dominance of period four lattice deformations has been obtained from self consistent computations\cite{Clay03}. A uniform deformation, leading to global size change, can also appear; this would produce a uniform shift in $t$ and $V$, which is inessential to our present analysis. \\
Before analizing the case $g_V>0$, we summarize all previous numerical results in the $1/4$  filled case with $g_V=0$.\\
When $g_P=g_H=0$ and $U\rightarrow\infty$ there exists a critical value of $V$ ($V_c=2\,t$) for the appearance of a $4k_f$ Charge-Density-Wave if $V>V_c$ Fig.\ \ref{pattern}(c). For $V<V_c$ the extended Hubbard model corresponds to a Luttinger liquid (LL) with no charge order. For finite $U$, the value of $V_c$ has been calculated  withing strong coupling perturbation theory\cite{PencMila94} and numerical methods\cite{Lin93}.\\
For small positive $g_P$ and $g_H$, and  $V<Vc$, the ground state has a dimerized $4k_F$ Bond-Order-Wave (BOW) with uniform site charges Fig.\ \ref{pattern}(a). If $g_P$ is sufficiently large ($g_H$ fixed), there is a superposition of the $2k_F$ and $4k_F$ BOWs accompanied by the $\bullet-\bullet-{\small\text{o}}-{\small\text{o}}$ charge order. This state correspond to the Bond-Charge-Density-Wave state (BCDW)  Fig.\ \ref{pattern}(b). On the other hand, at large $g_H$ and small $g_P$, a $4k_F$ CDW is found, with uniform bonds Fig.\ \ref{pattern}(c). If we increase $g_P$, the ground state becomes the $4k_F$ CDW-$2k_F$ SP, where the  charge order $\bullet-\bullet-{\small\text{o}}-{\small\text{o}}$, is accompanied by a $2k_F$ lattice distortion Fig.\ \ref{pattern}(d). For $V>Vc$, the $4k_F$ BOW phase is diminished due to the strong Coulomb repulsion $V$ and phases with charge-order dominate the phase diagram. In this work we present new  results for a $1/4$ filled band including the electron-phonon interaction (Peierls and Holstein couplings) and lattice deformations in the nearest-neighbor Coulomb interaction $V$ that has
not previously been studied. We show that this term may change substantially the phase diagram for large $V$.
%
\section{Bosonization approach and Semiclassical analysis}
In order to analize semi-quantitatively the low energy
properties of the model given by Eq.\ (\ref{H_total}), we use the Abelian
bosonization method which is generally powerful for the
description of one-dimensional chains (for further  details see Ref.\cite{Giamarchi04,CabraPujol04,Senechal99,Daniel01,Marion09}). Dimensionless parameters, are used below. They are introduced using $t$ as the energy scale as follows:
$g_P\rightarrow\tilde{g}_P=(t/K_P)^{1/2}g_P$, $g_H\rightarrow\tilde{g}_H=(t/K_H)^{1/2}g_H$,
$g_V\rightarrow\tilde{g}_V=(t/K_P)^{1/2}g_V$,
$\delta_{P,s}\rightarrow\tilde{\delta}_{P,s}=(K_P/t)^{1/2}\delta_{P,s}$ and
$\delta_{H,s}\rightarrow\tilde{\delta}_{H,s}=(K_H/t)^{1/2}\delta_{H,s}$ ($s=d,t$). To obtain
the corresponding low-energy, we write the fermion operator as\cite{Daniel01}
\begin{eqnarray}
c_{x,\sigma} \rightarrow \psi_{\sigma}(x) & \sim &e^{i k_{F,\sigma} x}~\psi_{L,\sigma}(x)+ e^{-i k_{F,\sigma} x}\,\psi_{R,\sigma}(x)\, + \ldots\nonumber \\
& = & e^{i k_{F,\sigma} x}~e^{-i \sqrt{4 \pi} \phi_{L,\sigma}(x)}\nonumber\\
&&+ e^{-i k_{F,\sigma} x}~e^{i \sqrt{4 \pi} \phi_{R,\sigma}(x)}
~ + \ldots ~,
\label{cfer}
\end{eqnarray}
where $k_{F,\sigma}$ are the Fermi momenta for up and down
spin electrons and $\phi_{R,L,\sigma}$ are the chiral components of two bosonic
fields, introduced as usual in order to bosonize the spin up and
down chiral fermion operators $\psi_{R,L,\sigma}$. The dots stand for higher order terms\cite{Daniel01}. They take into account the corrections arising from the curvature of the dispersion relation
due to the Coulomb interaction. For non-zero Hubbard repulsion $U$ and $V$, the low
energy effective Hamiltonian corresponding to (\ref{H_total}) written in terms of the bosonic fields $\phi_\uparrow$ and $\phi_\downarrow$ has a complicated form, mixing up and down degrees of freedom. We define $\phi_\sigma=\phi_{R,\sigma}+\phi_{L,\sigma}$ and introduce linear conbinations of $\phi_\sigma$ to describe the charge and spin  degrees of freedom,
\begin{eqnarray}
\phi_c = {1 \over \sqrt{2}} \left( \phi_\uparrow + \phi_\downarrow \right) ~, ~~
\phi_s = {1 \over \sqrt{2}}
\left( \phi_\uparrow - \phi_\downarrow \right) ~,
\end{eqnarray}

Then, we can rewrite $H=\int_0^L dx\,\mathcal{H}$
as (at $h=0$):
\begin{eqnarray}
\mathcal{H}&=&\mathcal{H}_{ph}+\mathcal{H}_{free}+V_{eff}
\label{TerminosH}
\end{eqnarray}
where
\begin{eqnarray}
\mathcal{H}_{ph}&=& \frac{1}{2}(\frac{1}{2}\,\tilde{\delta}_{P,t}^2+\,\tilde{\delta}_{P,d}^2)
+ \frac{1}{2}(\frac{1}{2}\,\tilde{\delta}_{H,t}^2+\,\tilde{\delta}_{H,d}^2)\nonumber\\
\mathcal{H}_{free}&=&\frac{v_c}{4\pi}
\left[
 \frac{1}{K_c}(\partial_x \phi_c )^2+ K_\rho(\partial_x \theta_c )^2
 \right]\nonumber\\
&&+
\frac{v_s}{4\pi}
\left[
 \frac{1}{K_s}(\partial_x \phi_s )^2
+ K_s(\partial_x \theta_s )^2
 \right]\nonumber\\
\mathcal{V}_{eff}&=&\mathcal{V}_{c}+\mathcal{V}_{s}+\mathcal{V}_{int}
\label{Hfree}
\end{eqnarray}
\begin{widetext}
\begin{eqnarray}
\mathcal{V}_{c}&=&\tilde{g}_{1/4}\, \cos 4\phi_c
+\lambda_1\,  \tilde{g}_{P} \,\tilde{\delta}_{P,d}
\sin 2 \phi_c+ \lambda_2\,
\tilde{g}_{H} \,\tilde{\delta}_{H,d}
 \cos 2\phi_c+\lambda_3\,\tilde{g}_{V}\,\tilde{\delta}_{P,d}\,\cos 2\phi_c\nonumber\\
\mathcal{V}_{\sigma}&=&\tilde{g}_s\, \cos 2\phi_s\nonumber\\
\mathcal{V}_{int}&=&-\tilde{g}_{P}\,\tilde{\delta}_{P,t}\cos\left(\phi_c - \xi  \right)\cos \phi_s
 - \tilde{g}_{H} \,\tilde{\delta}_{H,t}
  \sin \left(\phi_c +\zeta \right)
  \cos \phi_s-\lambda_4\,\tilde{g}_{V}\,\tilde{\delta}_{P,d}\,\cos 2\phi_c\,\cos 2\phi_s
 \nonumber \\
 &&+\lambda_5\,\tilde{g}_{V}\,\tilde{\delta}_{P,t}\,\cos \phi_s\sin\left(3\,\phi_c +
\xi-\frac{\pi}{4} \right)-\lambda_6\,\tilde{g}_{V}\,\tilde{\delta}_{P,t}\,\cos \phi_s\sin\left(\phi_c -\xi+\frac{\pi}{4} \right)
 \qquad
\label{Veff}
\end{eqnarray}
\end{widetext}
plus several vertex operators that are kept only when they are commensurate (non-oscillating in space)
and constitute relevant perturbations to the Gaussian conformal field theory. In  Eqs.\ (\ref{Hfree})
and (\ref{Veff})  $\phi_c$ and $\phi_s$ are the charge and spin fields respectively, $\theta_{c,s}$ are the dual  fields defined by $\partial_x\theta_{c,s}=\partial_t\phi_{c,s}$ and the phases $\xi$ and $\zeta$ determine the spatial patterns of the tetramerization (see Eq.\ (\ref{deformations})). The parameters
$v_{c}$ and $v_{s}$ are the velocity of the charge and spin excitations, and $K_{c}$ and $K_{s}$ are the corresponding Tomonaga-Luttinger parameters. From the perturbative calculation presented in \cite{Yoshioka00} the coupling constants are given by $\tilde{g}_{1/4}\varpropto U^2(U-4\,V)$ and $\tilde{g}_{s}\varpropto U-\alpha\,U(U-2\,V)$ where $\alpha$ is some numerical constant. Besides,
\begin{eqnarray}
\tilde{g}_{1/4}&\varpropto& U^2(U-4\,V)\nn\\
\tilde{g}_{s}&\varpropto& U-\alpha\,U(U-2\,V)\nn\\
\lambda_1&\varpropto&U\nn\\
\lambda_2&\varpropto&U\nn\\
\lambda_3&\varpropto& (U-cte)V\nn\\
\lambda_4&\varpropto& V\nn\\
\lambda_5&\varpropto& U\,V\nn\\
\lambda_6&\varpropto& (U-cte')V.
\label{acoplamientos}
\end{eqnarray}

The Eq.\ (\ref{Veff}) is the bosonic self-interaction potential defining a Multi-sine Gordon theory. Extensive analysis of the competition between different harmonics in multi-frequency sine-Gordon theories
have been performed in \cite{DelfinoMussardo98,Gogolin00,Bajnok01}, mainly focused on the
double sine-Gordon model. The three-frequency case has also been recently discussed in \cite{Toth04,Rosales07A,Rosales07B}. For our purposes it will be enough to perform a semiclassical treatment, as detailed in the next Section.

In order to perform the semiclassical analysis of the
ground state and present a schematic phase diagram, we use the bare coefficients
in Eq.\ (\ref{Veff}) and we assume the qualitative phenomenological dependence of the
coupling constants $\tilde{g}_{1/4}$, $\tilde{g}_{s}$ and $\lambda's$ on the microscopic
parameters as in \ref{acoplamientos}.

The aim of the present work is to search for the possibility
of elastic deformations that lower the energy with respect to the homogeneous non-deformed
case. The simplest analysis of the effective theory Eq.\ (\ref{TerminosH}), (\ref{Hfree}) and (\ref{Veff}), which has proven to be useful in related cases
\cite{Rosales07A,Rosales07B,Lecheminant04,Hida05}, consists in treating the
self-interaction terms in Eq.\ (\ref{Veff}) as a classical potential
to be evaluated in constant field configurations. Within this approximation the energy per site depends on four configuration parameters,
\begin{eqnarray}
\epsilon(\phi_c,\phi_s,\xi,\zeta)&=&\mathcal{V}_{eff}(\phi_c,\phi_s,\xi,\zeta)
\end{eqnarray}
so that the minima can be found analytically. In the previous expression, $\phi_c$ and $\phi_s$ are the charge and spin fields, and  the phases $\xi$ and $\zeta$ determine the spatial patterns of the tetramerization.

Our results on coupled Holstein-Peierls-Hubbard chain are summarized in Fig.\ \ref{phases}. First, we consider the case $\tilde{g}_V=0$. From the perturbative calculation, the coupling $\tilde{g}_{1/4}$ is positive  for small $V$ but becomes negative for large $V$. When $\tilde{g}_{1/4}>0$,
the phase variable $\phi_c$ is to be fixed at $\phi_c=\pi/4$, while for $\tilde{g}_{1/4}<0$, $\phi_c=0$.

Let us consider finite values of $\tilde{g}_P$ and $\tilde{g}_H$. In Fig.\ \ref{phases}(a)
we have set $\tilde{g}_{1/4}>0$ (or $U >4\,V$), and we increase $\tilde{g}_P$
from zero (for fixed small $\tilde{g}_{H}$). Unlike what has been found numerically in \cite{Clay03} (where no distortion patterns were favored, probably due to finite size effects), the phases
$(\phi_c,\phi_s)$ are locked in $(\frac{\pi}{4},\frac{\pi}{2})$ (independently of $\xi,\zeta$). In this case, the ground state becomes the $4k_F$ BOW while the site charge density remains uniform ($\langle n_i\rangle=1/2$), the
bond-order is now inhomogeneous and has the form shown in Fig.\ \ref{pattern}(a).
For $\tilde{g}_P$ larger than a critical values $\tilde{g}_P \agt \tilde{g}_P^c(\tilde{g}_H)$, we have $(\xi,\zeta,\phi_c,\phi_s)=
(\frac{\pi}{4},\frac{\pi}{4},\frac{\pi}{4},\pi)$ and the BCDW depicted in Fig.\ \ref{pattern}(b)
becomes the ground state. Hence the bond-order pattern now is $t(1+\alpha)$, $t(1-\alpha')$,
$t(1+\alpha)$, $t(1+\alpha'')$ and the charge order follows the ${\small\text{o}}-\bullet-\bullet-{\small\text{o}}$
pattern. Now consider the situation when $\tilde{g}_H$ reaches a critical value $\tilde{g}^c_H(\tilde{g}_P)$. In that case, $(\phi_c,\phi_s)=(0,\frac{\pi}{2})$ (independently of $\xi,\zeta$) and the $4k_F$ CDW with
${\bullet-\small\text{o}}-\bullet-{\small\text{o}}$ charge order and uniform bond order becomes
the ground state Fig.\ \ref{pattern}(c). The SP distortion, which corresponds to the  $4k_F$ CDW-$2k_F$ SP state (shown in Fig.\ \ref{pattern}(d) occurs only for large values of $\tilde{g}_H$ and $\tilde{g}_P\approx 0.8$. In this case $(\xi,\zeta,\phi_\rho,\phi_\sigma)=(\frac{\pi}{2},\pi,0,\pi)$.

 \begin{figure}[htbp]
  \centering
  \includegraphics[width=0.46\textwidth]{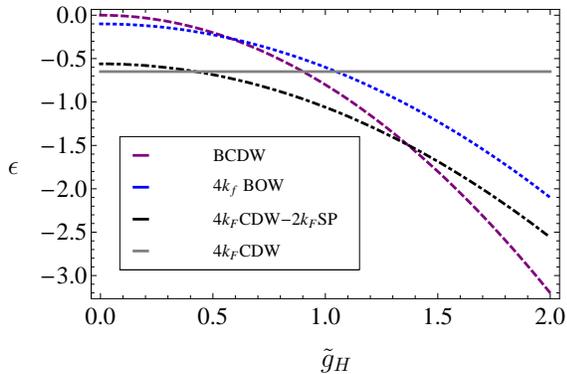}
      \caption{Semiclassical energies for the minima in terms of $\tilde{g}_P$
for $\tilde{g}_{1/4}=-1.5$, $\tilde{g}_H=2.2$ and $\tilde{g}_V=0$.}
     \label{EvsgP}
 \end{figure}
 \begin{figure}[htbp]
  \centering
  \includegraphics[width=0.52\textwidth]{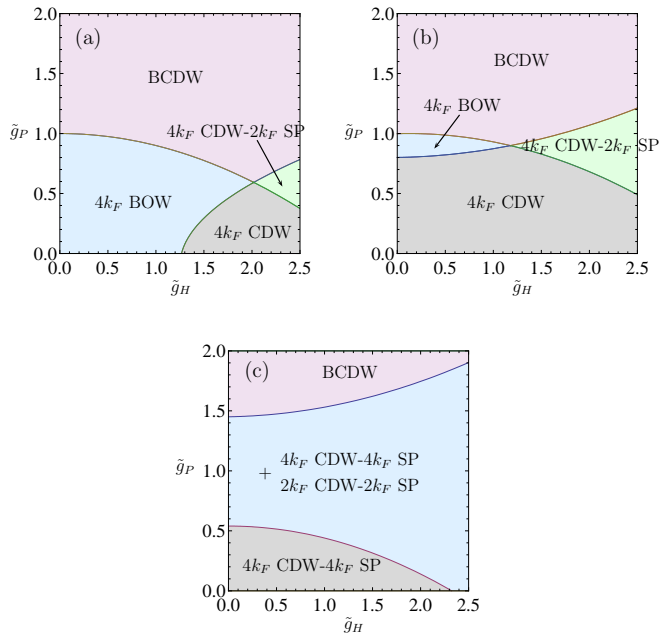}
     \caption{Schematic phase diagram in the plane $\tilde{g}_P$ vs $\tilde{g}_H$ for
different values of $\tilde{g}_V$. $(a)$ $\tilde{g}_{1/4}=1$, $\tilde{g}_V=0$; $(b)$
$\tilde{g}_{1/4}=-3$, $\tilde{g}_V=0$; $(d)$ $\tilde{g}_{1/4}=-3$, $\tilde{g}_V=0.2$. The boundaries of the  phases result from
level crossings as shown in the Fig.\ (\ref{EvsgP}). The case with $\tilde{g}_{1/4}>0$ and $\tilde{g}_V>0$ is similar to the case $(a)$ and we don't show here. The effect of the Peierls deformation in the nearest-neighbor Coulomb repulsion is more important  for large $V$, that is, $\tilde{g}_{1/4}<0$ .}
     \label{phases}
 \end{figure}
In Fig.\ \ref{phases}(b) we can see several important trends when  $\tilde{g}_{1/4}<0$ (or $V> U/4$).
First, the size of the  $4k_F$ BOW region decreases considerably with decreasing $\tilde{g}_{1/4}<0$,
showing that it takes stronger e-ph coupling to overcome the ``natural'' tendency towards
${\bullet-\small\text{o}}-\bullet-{\small\text{o}}$ charge order for small $\tilde{g}_{P}$. For
the same reason, the sizes of the $4k_F$ CDW and $4k_F$ CDW-$2k_F$ SP regions increase considerably
when $\tilde{g}_{1/4}$ is negative. Next we examine the effect of $\tilde{g}_{V}$. When $\tilde{g}_{V}=0.2$
and $\tilde{g}_{1/4}<0$, we can see that the $4k_F$ BOW region ``disappears'' and for $\tilde{g}_{P}\lesssim 1.5$
the new phases   $4k_F$ CDW-$4k_F$ SP and $4k_F$ CDW-$4k_F$ SP + $2k_F$ CDW-$2k_F$ SP dominate. Unlike the
previous case with $\tilde{g}_{V}=0$, in the these  phases, there is now a small dimerization in the bonds (Fig.\ \ref{pattern}(e) and Fig.\ \ref{pattern}(f)).\\
\section{Summary and Discussions}
We have investigated the electronic ground-state phase diagram of the one-dimensional Hubbard model at quarter filling with on-site ($U$) and nearest-neighbor ($V$) Coulomb repulsion coupled to adiabatic phonons that includes Holstein and Peierls interactions. In addition to the conventional terms, we have included the effects of the Peierls deformation in the nearest-neighbor Coulomb interaction $V$ that may be important when the Coulomb interaction between two sites $r_i$ and  $r_j$ is exponentially short ranged, {\it i.e.} $\exp[-\alpha|r_i-r_j|]/r^\xi$ ($\alpha,\xi>0$). We have used Abelian bosonization  to analyze the low energy properties of the model and we have performed a semiclassical analysis of the bosonized effective theory to identify the possible phases of the ground state. We have found that several charge orders combined  with lattice deformations describe the ground states of the system depending on the microscopic parameters $U,V$ and the spin-phonon couplings $\tilde{g}_P$, $\tilde{g}_H$ and $\tilde{g}_V$. In each of these phases a non trivial elastic deformation is favored, grouping together blocks of two or four sites, while the charge sector adopts different charge orders. \\
A detailed analysis for particular values of $\tilde{g}_V$, shows the following  phases at zero temperature:\\
\begin{itemize}
 \item For $U>V$ ($\tilde{g}_{1/4}>0$), $\tilde{g}_V=0$ and for small and nonzero $\tilde{g}_P$ and $\tilde{g}_H$, the ground state has a dimerized $4k_F$ Bond-Order-wave (BOW) with uniform site charges Fig.\ \ref{pattern}(a). If $\tilde{g}_P$ is sufficiently large ($\tilde{g}_H$ fixed), there is a  second dimerization of the dimerized lattice, leading to the BCDW state (a superposition of the $2k_F$ and $4k_F$ BOWs), accompanied by the $\bullet-\bullet-{\small\text{o}}-{\small\text{o}}$ charge order Fig.\ \ref{pattern}(b). On the other hand, at large $\tilde{g}_H$ and small $\tilde{g}_P$, a $4k_F$ CDW is found, with uniform bonds Fig.\ \ref{pattern}(c). If we increase $\tilde{g}_P$, the ground state becomes the $4k_F$ CDW-$2k_F$ SP, where the  charge order $\bullet-\bullet-{\small\text{o}}-{\small\text{o}}$, is accompanied by a $2k_F$ lattice distortion Fig.\ \ref{pattern}(d).\\
\item For $V>U$ ($\tilde{g}_{1/4}<0$), the $4k_F$ BOW phase is diminished due to the strong Coulomb repulsion $V$ and phases with charge-order dominate the phase diagram Fig.\ (\ref{phases}).\\
\item For $V>U$ ($\tilde{g}_{1/4}<0$), $\tilde{g}_V>0$, the $4k_F$ BOW disappears because the Coulomb interaction between nearest-neighbor is increased due the Peierls deformation. The CDW's phases   has a small  contribution of $4k_F$ BOWs combined with the original $2k_F$, giving rise to two new phases: $4k_F$ CDW-$4k_F$ SP and $4k_F$ CDW-$4k_F$ SP + $2k_F$ CDW-$2k_F$ SP
\end{itemize}

In conclusion, we have argued that the coexistence of charge order
and lattice distortions is dominated by $4k_F$ BOW, BCDW for $\tilde{g}_{1/4}>0$. On the one
hand, when $\tilde{g}_{1/4}<0$, the $4k_F$ CDW and $4k_F$ CDW-$2k_F$SP dominate if
$\tilde{g}_{P}<\tilde{g}_{P}^c$,\,  Fig.\ \ref{phases}(b); for $\tilde{g}_{P}>\tilde{g}_{P}^c$  the
BCDW  phase dominate. On the other hand, when the phonon-lattice coupling, $\tilde{g}_V$ is taken into
account (only for $\tilde{g}_{1/4}<0$ because is necessary $V>U$), two new phases that we
named $4k_F$ CDW-$4k_F$ SP and  $4k_F$ CDW-$4k_F$ SP+$2k_F$ CDW-$2k_F$ SP (Fig.\ \ref{pattern}(e) and Fig.\ \ref{pattern}(f) respectively) dominate the phase
diagram. These phases differ from the original $4k_F$ CDW and $4k_F$ CDW-SP, in  a small dimerization
in the bonds, $\tilde{\delta}_{P,d}$ and a small contribution of $2k_F$ CDW in the last one. \\
Numerical analysis of the present system including Peierls deformations in the nearest-neighbor coulomb interaction would be helpful in order to determine the phase boundaries quantitatively. We hope that our predictions will stimulate further X-ray or neutron scattering experiments on compounds like Bechgaard salts. They should be able to distinguish the different distortion patters predicted  here. Since the effects of the Peirels deformations in the nearest-neighbor coulomb interaction is proportional to $V$ (see Eq.\ (\ref{H_total1})), we expect that the effects of this term will be more important in compounds where $V$ is larger.
\section{Acknowledgments}
We  thank A.\ Honecker, G.I.\ Japaridze, C.A.\ Lamas, M.\ Moliner, D.\ Poilblanc,
P.\ Pujol and  G.L.\ Rossini for helpful discussions.\\
This work was partially supported by the ESF grant INSTANS, ECOS-Sud
Argentina-France collaboration (Grant No A04E03), PICS CNRS-Conicet
(Grant No. 18294), PICT ANCYPT (Grant No 20350), and PIP CONICET
(Grant No 5037).



\begin{thebibliography}{99}
%
\bibitem{Salts}
For reviews on Bechgaard salts see {\it e.g.} D.\ J\'erome and H.J.\ Schulz, Adv.\ Phys. \textbf{31}, 299 (1982); D.\ Jérome, {\it Organic superconductors: From (TMTSF)$_2$PF$_6$ to fullerenes} (Marcel Dekker, New York, 1994), 405. T.\ Ishiguro and K.\ Yamaji, Springer\ Series\ Solid\ State Vol. \textbf{88}, Springer-Verlag, Berlin (1990). T.\ Ishiguro, K.\ Yamaji and G.\ Saito, {\it Organics Superconductors}, Springer-Verlag, New York, (1998).
%
%
\bibitem{Hiraki98}
K.\ Hiraki and K.\ Kanoda,
   Phys.\ Rev.\ Lett. \textbf{80}, 4737 (1998).
%
\bibitem{Mazumdar99B}
S.\ Mazumdar, D.\ Campbell, R.T.\ Clay and S.\ Ramasesha, Phys.\ Rev.\ Lett. \textbf{82}, 2411 (1999).
%
\bibitem{Hiraki99}
K.\ Hiraki and K.\ Kanoda, Phys.\ Rev.\ Lett. \textbf{82}, 2412 (1999).
%
\bibitem{Biskup}
N.\ Biskup, J.A.A.J.\ Perenboom, J.S.\ Brooks, and J.S.\ Qualls, Solid\ State\ Comm. \textbf{107}, 503 (1998).
%
\bibitem{Chow}
D.S.\ Chow, F.\ Zamborszky, B.\ Alavi, D.J.\ Tantillo, A.\ Baur, C.A.\ Merlic, and S.E.\ Brown, Phys.\ Rev.\ Lett. \textbf{85}, 1698 (2000).
%
\bibitem{Nad00A}
F.\ Nad, P.\ Monceau, C.\ Carcel and J.M.\ Fabre, J.\ Phys.\ Cond.\ Matt. \textbf{12}, L435 (2000).
%
\bibitem{Nad00B}
F.\ Nad, P.\ Monceau, C.\ Carcel, and J.M.\ Fabre, Phys.\ Rev.\ B \textbf{62}, 1753 (2000).
%
\bibitem{Monceau}
P.\ Monceau, F.Y.\ Nad, and S.\ Brazovskii, Phys.\ Rev.\ Lett. \textbf{86}, 4080 (2001).
%
\bibitem{Miyagawa}
K.\ Miyagawa, A.\ Kawamoto and K.\ Kanoda, Phys.\ Rev.\ B \textbf{62}, R7679 (2000).


\bibitem{Ung}
K.C.\ Ung, S.\ Mazumdar and D.\ Toussaint, Phys.\ Rev.\ Lett. \textbf{73}, 2603 (1994).
%
\bibitem{Mazumdar99A}
S.\ Mazumdar, S.\ Ramasesha, R.T.\ Clay and D.K.\ Campbell, Phys.\ Rev.\ Lett. \textbf{82}, 1522 (1999).
%
\bibitem{Kobayashi97}
N.\ Kobayashi and M.\ Ogata, J.\ Phys.\ Soc.\ Jpn. \textbf{66}, 3356 (1997).
%
\bibitem{Kobayashi98}
N.\ Kobayashi, M.\ Ogata  and K.\ Yonemitsu, J.\ Phys.\ Soc.\ Jpn. \textbf{67}, 1098 (1998).
%
\bibitem{Riera99}
J.\ Riera and D.\ Poilblanc, Phys.\ Rev.\ B \textbf{59}, 2667 (1999).
%
\bibitem{Riera00}
 J.\ Riera and D.\ Poilblanc, Phys.\ Rev.\ B \textbf{62}, R16243 (2000).
%
\bibitem{Seo97}
H.\ Seo and H.\ Fukuyama, J.\ Phys.\ Soc.\ Jpn. \textbf{66}, 1249 (1997).
%
\bibitem{Seo00}
H.\ Seo, J.\ Phys.\ Soc.\ Jpn. \textbf{69}, 805  (2000).
%
\bibitem{Clay03}
R.T.\ Clay, S.\ Mazumdar and D.K.\ Campbell, Phys.\ Rev.\ B \textbf{67}, 115121 (2003).
%
\bibitem{Daniel01}
D.C.\ Cabra, A.\ De Martino, A.\ Honecker, P.\ Pujol,\ and\ P.\ Simon,
Phys.\ Rev. B \textbf{63}, 094406 (2001).
%
\bibitem{Marion09}
For a recent review on the Hubbard chain coupled to adiabatic phonons, see for example chapter IV in
http://eprints-scd-ulp.u-strasbg.fr:8080/1100/01/MOLINER\_Marion\_2009.pdf.
%
%
\bibitem{PencMila94}
K.\ Penc and F.\ Mila, Phys.\ Rev.\ B \textbf{49}, 9670 (1994).
%
\bibitem{Lin93}
H.Q.\ Lin, {\it et al}, Proceedings of the 1993 NATO ARW on {\it The
  Physics and Mathematical Physics of the Hubbard Model}, edited by D.\ Baeriswyl {\it et al}, Plenum, New York (1995).
%
\bibitem{Giamarchi04}
T.\ Giamarchi, {\it Quantum Physics in One Dimension}, Oxford\ University\ Press, Oxford (2004).
%
\bibitem{CabraPujol04}
See for instance D.C.\ Cabra and P.\ Pujol, in {\it Quantum Magnetism}, Lect.\ Notes\ in \ Phys. \textbf{645} 253 (2004).
%
\bibitem{Senechal99}
D.\ S\`en\'echal, {\it An introduction to bosonization},
arXiv:\ cond-mat 9908262v1 (1999) or in the book D.\ S\`en\'echal, A.-M.\ Tremblay and C.\ Bourbonnais {\it Theoretical Methods for Strongly Correlated Electrons},  Springer, 1 edition (October 1, 2003).
%
\bibitem{Yoshioka00}
H.\ Yoshioka, M.\ Tsuchiizu and Y.\ Suzumura,
J.\ Phys.\ Soc.\ Jap. \textbf{69}, 651 (2000).
%
\bibitem{DelfinoMussardo98}
G.\ Delfino and G.\ Mussardo, Nucl.\ Phys.\ B \textbf{516}, 675 (1998).
%
\bibitem{Gogolin00}
M.\ Fabrizio, A.O.\ Gogolin, and A.A.\ Nersesyan, Nucl.\ Phys.\ B \textbf{580}, 647 (2000).
%
\bibitem{Bajnok01}
Z.\ Bajnok, L.\ Palla, G.\ Takacs, and F.\ Wagner, Nucl.\ Phys.\ B \textbf{601}, 503 (2001).
%
\bibitem{Toth04}
G.Z.\ T\'oth, J.\ Phys.\ A \textbf{37}, 9631 (2004).
%
\bibitem{Rosales07A}
H.D.\ Rosales, D.C.\ Cabra, M.D.\ Grynberg, G.L.\ Rossini, and T.\ Vekua, Phys.\ Rev.\ B \textbf{75},
174446 (2007).
%
\bibitem{Rosales07B}
H.D.\ Rosales and G.L.\ Rossini, Phys.\ Rev.\ B \textbf{76}, 224404 (2007).
%
\bibitem{Lecheminant04}
P.\ Lecheminant and E.\ Orignac, Phys.\ Rev.\  B \textbf{69}, 174409 (2004).
%
\bibitem{Hida05}
K.\ Hida and I.\ Affleck, J.\ Phys.\ Soc.\ Jpn.\ \textbf{74}, 1849 (2005).
%
\end{thebibliography}
\end{document}